\documentclass[aps, twocolumn, a4paper, superscriptaddress]{revtex4}

\usepackage{graphicx}
\usepackage{float}
\usepackage{color}
\usepackage{amsmath, amssymb}
\usepackage{lmodern}
\usepackage[T1]{fontenc}
\usepackage[utf8]{inputenc}
\usepackage[english]{babel}

\def\lsim{\mathrel{\rlap{\lower4pt\hbox{\hskip1pt$\sim$}}
    \raise1pt\hbox{$<$}}}                % less than or approx. symbol
\def\gsim{\mathrel{\rlap{\lower4pt\hbox{\hskip1pt$\sim$}}
    \raise1pt\hbox{$>$}}}                % greater than or approx. symbol

\begin{document}

\title{Predominance of the weakest species in Lotka-Volterra and May-Leonard implementations of the rock-paper-scissors model}

\author{P.P. Avelino}
\affiliation{Instituto de Astrof\'{\i}sica e Ci\^encias do Espa{\c
c}o, Universidade do Porto, CAUP, Rua das Estrelas, PT4150-762 Porto,
Portugal}
\affiliation{Departamento de F\'{\i}sica e Astronomia, Faculdade de
Ci\^encias, Universidade do Porto, Rua do Campo Alegre 687, PT4169-007
Porto, Portugal}
\author{B.F. de Oliveira}
\affiliation{Departamento de Física,
Universidade Estadual de Maringá, 
Av. Colombo 5790, 87020-900 Maringá, PR, Brazil}
\author{R.S. Trintin}
\affiliation{Departamento de Física,
Universidade Estadual de Maringá, 
Av. Colombo 5790, 87020-900 Maringá, PR, Brazil}

\begin{abstract}
We revisit the problem of the predominance of the `weakest' species in the context of Lotka-Volterra and May-Leonard implementations of a spatial stochastic rock-paper-scissors model in which one of the species has its predation probability reduced by $0<{\mathcal P}_w<1$. We show that, despite the different population dynamics and spatial patterns, these two implementations lead to qualitatively similar results for the late time values of the relative abundances of the three species (as a function of ${\mathcal P}_w$), as long as the simulation lattices are sufficiently large for coexistence to prevail --- the `weakest' species generally having an advantage over the others (specially over its predator). However, for smaller simulation lattices, we find that the relatively large oscillations at the initial stages of simulations with random initial conditions may result in a significant dependence of the probability of species survival on the lattice size and total simulation time. 
\end{abstract}

\maketitle

\section{Introduction \label{sec1}}

Non-hierarchical competition interactions have been shown to play a crucial role in the preservation of coexistence. The classical rock-paper-scissors (RPS) model \cite{2002-Kerr-N-418-171,Reichenbach-N-448-1046}, the simplest cyclic predator-prey model, describes the dynamics  of three different species subject to interspecific competition (see \cite{1920PNAS....6..410L,1926Natur.118..558V,May-Leonard}   for the pioneer work by Lotka and Volterra, and May and Leonard). It allows for the stable coexistence of all three species, and  successfully reproduces some of the main dynamical features observed in simple biological systems with cyclic selection interactions \cite{lizards,2002-Kerr-N-418-171,bacteria}.

Simulations of the spatial RPS model are usually performed on a square lattice (see \cite{2004-Szabo-JPAMG-31-2599, 2009-Zhang-PRE-79-062901, 2014-Laird-Oikos-123-472, 2014-Rulquin-PRE-89-032133}, however, for other lattice configurations) and consider nearest-neighbor cyclic predator-prey interactions. In a Lotka-Volterra implementation of the RPS model, each site is occupied by a single individual of one of the three species, and there is a conservation law for the total number of individuals. On the other hand, in a May-Leonard implementation each site may either be occupied by a single individual or left empty, and the total number of individuals is, in general, not conserved \cite{2008-Peltomaki-PRE-78-031906}.

Generalizations of the RPS model involving additional species and interactions, have also been investigated in recent years \cite{2008-Szabo-PRE-77-041919, 2011-Allesina-PNAS-108-5638, 2012-Avelino-PRE-86-031119, 2012-Avelino-PRE-86-036112,  2012-Li-PA-391-125, 2012-Roman-JSMTE-2012-p07014, 2013-Lutz-JTB-317-286, 2013-Roman-PRE-87-032148, 2014-Cheng-SR-4-7486, 2014-Szolnoki-JRSI-11-0735, 2016-Kang-Entropy-18-284, 2016-Roman-JTB-403-10, 2017-Brown-PRE-96-012147, 2017-Park-SR-7-7465, 2017-Bazeia-EPL-119-58003, 2017-Souza-Filho-PRE-95-062411, 2018-Shadisadt-PRE-98-062105, 2019-Avelino-PRE-99-052310}. Complex dynamical spatial structures (such as spirals with an arbitrary number of arms \cite{2012-Avelino-PRE-86-036112, 2017-Bazeia-EPL-119-58003, 2019-Bazeia-PRE-99-052408}, domain interfaces, with or without  non-trivial internal dynamics \cite{2014-Avelino-PRE-89-042710}, and string networks, with or without junctions \cite{2014-Avelino-PLA-378-393, 2017-Avelino-PLA-381-1014}), diverse scaling laws \cite{2012-Avelino-PRE-86-036112, 2017-Brown-PRE-96-012147}, and phase transitions \cite{2001-Szabo-PRE-63-061904, 2004-Szabo-PRE-69-031911, 2004-Szolnoki-PRE-70-037102, 2007-Perc-PRE-75-052102, 2007-Szabo-PRE-76-051921, 2008-Szabo-PRE-77-011906, 2011-Szolnoki-PRE-84-046106, 2013-Vukov-PRE-88-022123, 2018-Bazeia-EPL-124-68001} have been shown to naturally emerge in some of these scenarios.

In most of these models the species may be characterized as having equal strength, with the survival probability being mainly dependent on initial conditions. However, there are other situations in which there is a competitive difference between species, such as in the case of a RPS model in which one of the species (often termed the  `weakest') has a reduced  predation probability ${\mathcal P}_w$. It has been shown in Refs.  \cite{2001-Frean-PRSLB-268-1323,2009PhRvL.102d8102B} that in a Lotka-Volterra implementation of this model, the `weakest' species tends to be the most abundant. These results have recently been challenged in Ref. \cite{2019-Menezes-EPL-126-18003}, with the authors claiming that some of the  model parameters have a significant impact on which species survives in a May-Leonard implementation.

In this paper we shall address the question of whether the predominance of the `weakest' species generally occurs in both Lotka-Volterra and May-Leonard implementations of the RPS model. The outline of this paper is as follows. In Sec. \ref{sec2} we start by considering a non-spatial RPS model in which one of the species has a reduced predation probability, discussing the properties of its stationary solutions in both Lotka-Volterra and May-Leonard implementations. In Sec. \ref{sec3} we describe these two implementations of the spatial stochastic RPS model and present the corresponding results. Special emphasis is given to the dependence of the survival probability on the size of the simulation lattice and total simulation time, and to the way the average densities of the three species depend on the reduced predation probability, parameterized by ${\mathcal P}_w$, for sufficiently large simulation lattices. Finally, we conclude in Sec. \ref{sec5}.

\section{Non-spatial RPS model \label{sec2}}

Let us start by considering Lotka-Volterra and May-Leonard implementations of the non-spatial RPS model.

\subsection{Lotka-Volterra \label{sub1}}

A Lotka-Volterra implementation of the RPS model considers three species with densities $\rho_1$, $\rho_2$ and $\rho_3$, such that $\rho_1+\rho_2+\rho_3=1$ (the total density is normalized to unity). At each timestep an individual of one of the species $i$ is selected at random and the predation interaction
\begin{equation}
i\ (i+1) \to i\ i\,, \nonumber
\end{equation}
with $i=1,...,3$, is performed with probability $p_i$. In this paper, modular arithmetic, where integers wrap around upon reaching $1$ or $3$, is assumed (the integers $i$ and $j$ represent the same species whenever $i=j \, {\rm mod} \, 3$, where $\rm mod$ denotes the modulo operation).

With an appropriate choice of time unit, the equations for the evolution of the densities of the different species may be written as
\begin{equation}
{\dot \rho_i}=p_i \, \rho_i \, \rho_{i+1}-p_{i-1} \, \rho_{i-1} \, \rho_i\,,
\label{lvns}
\end{equation}
where a dot represents a derivative with respect to time. Stationary solutions to Eq. (\ref{lvns}) satisfy the condition ${\dot \rho_i}=0$, and are, therefore, characterized by
\begin{equation}
\rho_{i+1}=\frac{p_{i-1}}{p_i}{\rho_{i-1}} \,, \qquad  \sum_{i=1}^3 \rho_i=1\,. \label{lvns1}
\end{equation}
Here, we shall be interested in the case where $p_1={\mathcal P}_w \, p$ and $p_2=p_3=p$, with $0<{\mathcal P}_w<1$, so that $\rho_1=\rho_2=\rho_3/{\mathcal P}_w$. Hence, Eq. (\ref{lvns1}) implies that the stationary solutions of Eq. (\ref{lvns}) are characterized by
\begin{equation}
\rho_1=\rho_2=\frac{1}{2+{\mathcal P}_w} \, \qquad \rho_3 = \frac{{\mathcal P}_w}{2+{\mathcal P}_w}\,,
\label{rho123LV}
\end{equation}
with $\rho_1 = \rho_2 > \rho_3$.

\subsection{May-Leonard \label{sub2}}

In a May-Leonard implementation of the RPS model the total density of individuals is no longer conserved. In this case, $\rho_0+\rho_1+\rho_2+\rho_3=1$ where, for uniformity of notation, $\rho_0$ shall be referred to as the density of empty sites  --- denoted by a `0' ---  even when considering a non-spatial RPS model. At each timestep an individual of one of the species $i$ is selected at random and an interaction is performed: either predation 
\begin{equation}
i\ (i+1) \to i\ 0\,, \nonumber
\end{equation}
with probability $p_i$, or reproduction
\begin{equation}
 i\ 0 \to ii\,, \nonumber
\end{equation} 
with probability $r$ (assumed to be the same for all the species) --- notice that predation has a different meaning in Lotka-Volterra and May-Leonard implementations of the RPS model. Again, with an appropriate choice of time unit, the equations for the evolution of the densities of the different species may be written as
\begin{equation}
{\dot \rho_i}=r \, \rho_i  \, \rho_{0}-p_{i-1} \, \rho_{i-1} \, \rho_i\,,
\label{mlnsi}
\end{equation}
while the evolution of the density of empty sites is given by
\begin{equation}
{\dot \rho_0}=-r  \, \rho_{0} \sum_{i=1}^3 \rho_i +\sum_{i=1}^3 p_{i-1} \, \rho_{i-1} \, \rho_i\,.
\label{mlns0}
\end{equation}
Stationary solutions to Eqs. (\ref{mlnsi}) and (\ref{mlns0}) satisfy the conditions ${\dot \rho_i}=0$ and ${\dot \rho_0}=0$, and are, therefore, characterized by
\begin{equation}
p_{i-1} \, \rho_{i-1}=r \, \rho_0 \,, \qquad \rho_0 + \sum_{i=1}^3 \rho_i=1\,.
\label{mlns1}
\end{equation}
Again, we shall be interested in the case where $p_1={\mathcal P}_w p$, with $p_2=p_3=p$, with $0<{\mathcal P}_w<1$. Equation (\ref{mlns1}) implies that the stationary solutions to Eqs. (\ref{mlnsi}) and (\ref{mlns0}) are characterized by 
\begin{eqnarray}
\rho_0 &=&\frac{1}{1+\frac{r}{p}\left(2+\frac{1}{{\mathcal P}_w}\right)}\,, \label{rho0123MLa}\\
\rho_1 &=&\frac{\frac{r}{p{\mathcal P}_w}}{1+\frac{r}{p}\left(2+\frac{1}{{\mathcal P}_w}\right)} \label{rho0123MLb}\,,\\
\rho_2=\rho_3 &=&\frac{\frac{r}{p}}{1+\frac{r}{p}\left(2+\frac{1}{{\mathcal P}_w}\right)}\,,
\label{rho0123MLc}
\end{eqnarray}
with $\rho_1=\rho_2/{\mathcal P}_w= \rho_3/{\mathcal P}_w$, so that $\rho_1 > \rho_2=\rho_3$.

Hence, we may conclude that in both Lotka-Volterra and May-Leonard implementations of the non-spatial RPS model the `weakest' species ($1$) has a competitive advantage. In a May-Leonard implementation the stationary density of individuals of the `weakest' species is larger than that of the other two species. On the other hand, in a Lotka-Volterra implementation the competitive advantage is less pronounced, since the stationary density of the `weakest' species is only larger than that of its predator (its prey having an equal density).

\begin{figure}[t]
	\centering
		\includegraphics{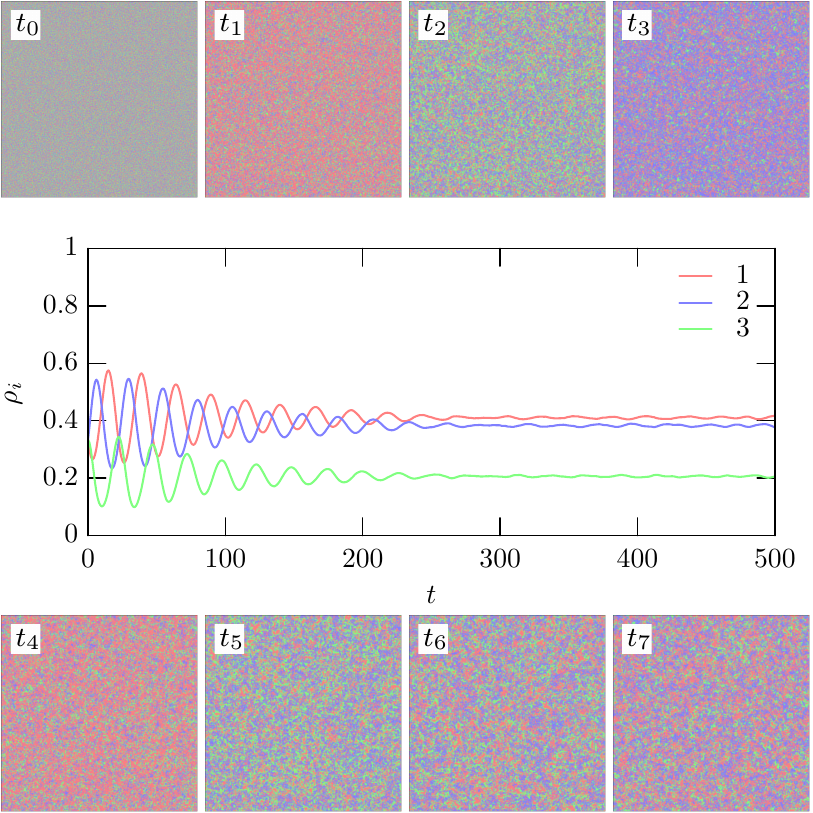}
	\caption{The upper and lower panels display snapshots of the spatial distribution of the different species on a  $1000^2$ lattice at $t_0 = 0$, $t_1 = 16$, $t_2 = 23$, $t_3= 30$, $t_4 = 40$, $t_5 = 48$, $t_6 = 98$, and $t_7 = 500$ for a single Lotka-Volterra realization of the spatial stochastic RPS model with random initial conditions (for $m=0.5$, $p=0.5$, and ${\mathcal P}_w=0.5$). The central panel shows the evolution of the density of the different species $\rho_i$ for the entire timespan of the simulation ($\rho_1 = \rho_2 = \rho_3 = 1/3$ at the initial time $t_0$). }
	\label{fig1}
\end{figure}

\section{Spatial rock-paper-scissors model \label{sec3}}

In this section we shall describe Lotka-Volterrra and May-Leonard implementations of the spatial RPS model which we shall investigate in the present paper. To this end, we shall consider a $N^2$ square lattice --- $N$ shall be referred to as its linear size --- with ${\mathcal N}$ sites and periodic boundary conditions. In a Lotka-Volterra implementation every site is occupied by a single individual of one of the three-species, while in a May-Leonard implementation there is also the possibility of a site being empty. The number of individuals of the species $i$ and the number of empty sites will be denoted by $I_i$ and $I_0$, respectively --- the density of individuals of the species $i$ and the density of empty sites shall be defined by $\rho_i=I_i/{\mathcal N}$ and $\rho_0 = I_0/{\mathcal N}$, respectively (note that $\rho_0=0$ in a Lotka-Volterra implementation). The possible interactions are the ones described in the case of the non-spatial RPS model, plus mobility
\begin{equation}
 i\ \odot \to \odot\ i\,, \nonumber
\end{equation}
where $\odot$ represents either an individual of any species or an empty site. 

%For the sake of definiteness, we shall take $m=0.5$, $p=0.25$ and $r=0.25$ in all the simulations {\color{blue} for May-Leonard and $m=0.5$ and $p=0.5$ for Lotka-Volterra, unless stated otherwise}. 

At every simulation step, the algorithm randomly picks an occupied site to be the active one, randomly selects one of its adjacent neighbour sites to be the passive one, and randomly chooses an interaction to be executed by the individual at the active position: predation, mobility or reproduction with probabilities $p$, $m$ and $r$, respectively ($r=0$ in a Lotka-Volterra implementation) --- except if stated otherwise, in this paper we use the von Neumann neighbourhood (or 4-neighbourhood) composed of a central cell (the active one) and its four non-diagonal adjacent cells. These three actions are repeated until a possible interaction is selected --- note that in both implementations of the RPS model the interaction cannot be carried out whenever predation is selected and the passive is not a prey of the active, while in a May-Leonard implementation the interaction is not completed also if reproduction is selected and the passive is not an empty site.

\begin{figure}[t]
	\centering
		\includegraphics{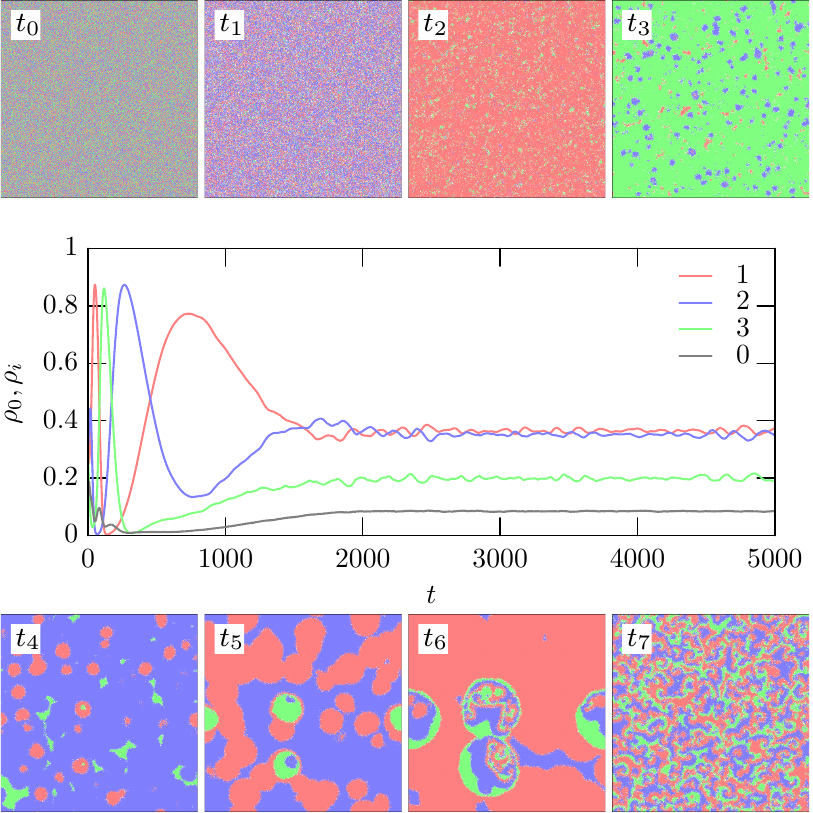}
	\caption{The upper and lower panels display snapshots of the spatial distribution of the different species on a  $1000^2$ lattice at $t_0 = 0$, $t_1 = 17$, $t_2 =
51$, $t_3 = 118$, $t_4 = 264$, $t_5 = 454$, $t_6 = 740$, and $t_7 = 5000$ for a single May-Leonard realization of the spatial stochastic RPS model with random initial conditions (for $m=0.5$, $p=0.25$, $m=0.25$ and ${\mathcal P}_w=0.5$). The central panel shows the density of the different species  and empty sites ($\rho_i$ and $\rho_0$, respectively) for the entire timespan of the simulation ($\rho_1 = \rho_2 = \rho_3 = 1/3$ at the initial time $t_0$).}
	\label{fig2}
\end{figure}

A generation time (our time unit) is defined as the time necessary for $\mathcal{N}$ successive interactions to be completed. 

\subsection{Results \label{sec4}}

Figures \ref{fig1} and \ref{fig2} show, respectively, the population network evolution in Lotka-Volterra and May-Leonard implementations of the spatial stochastic RPS model with random initial conditions --- each site being initially occupied by a randomly chosen single individual of any of the three species with a uniform discrete probability of $1/3$. 

In Fig. \ref{fig1} (Lotka-Volterra implementation) the upper and lower panels display snapshots of the spatial distribution of the different species on a $1000^2$ lattice at $t_0 = 0$, $t_1 = 16$, $t_2 = 23$, $t_3= 30$, $t_4 = 40$, $t_5 = 48$, $t_6 = 98$, and $t_7 = 500$, for a single Lotka-Volterra realization of the spatial stochastic RPS model with $m=0.5$, $p=0.5$, and ${\mathcal P}_w=0.5$ ---  species $1$, $2$, and $3$ are represented in red, blue and green, respectively. Notice the change in the overall color tone which takes place in the early stages of the simulation, associated to changes in the densities of the three species. Such oscillations are clearly visible in the central panel of Fig. \ref{fig1} which shows the evolution of the density $\rho_i$ of the different species for the entire timespan of the simulation --- the red, blue and green lines (from top to bottom, respectively) representing the densities of species 1, 2 and 3, respectively. Fig. \ref{fig1} shows that after short transient initial stage, with relatively large coherent oscillations, the densities of the three species quickly approach nearly constant values, with $\rho_1>\rho_2>\rho_3$. It reveals the predominance of the `weakest' species ($1$), specially over its predator ($3$).

\begin{figure}[t]
	\centering
		\includegraphics{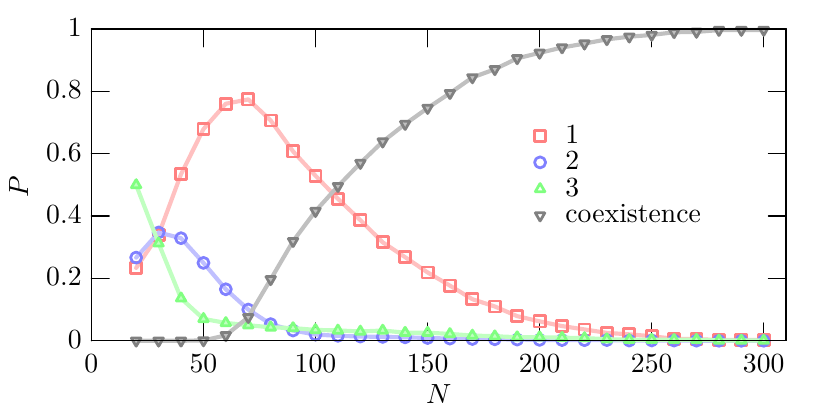}
	\caption{Probability of single species survival and coexistence as a function of the linear lattice size for a May-Leonard implementation of the spatial stochastic RPS model with $m=0.3$, $p=0.35$, $r=0.35$ and ${\mathcal P}_w=0.5$. Each point was estimated from $10^4$ simulations with a total simulation time equal to $2 \times 10^4$ generations, starting from random initial conditions with $\rho_1=\rho_2 = \rho_3 =
1/3$. The error bars are always much smaller than the size of the symbols.}
	\label{fig3}
\end{figure}

In Fig. \ref{fig2} (May-Leonard implementation) the upper and lower panels show snapshots of the spatial distribution of the different species on a $1000^2$ lattice at $t_0 = 0$, $t_1 = 17$, $t_2 =
51$, $t_3 = 118$, $t_4 = 264$, $t_5 = 454$, $t_6 = 740$, and $t_7 = 5000$, for a single May-Leonard realization of the spatial stochastic RPS model with $m=0.5$, $p=0.25$, $r=0.25$, and ${\mathcal P}_w=0.5$ --- species $1$, $2$, and $3$, and empty sites are represented in red, blue, green and white, respectively. The most prominent feature in the snapshots shown in Fig. \ref{fig2} is the presence of distinctive compact spatial domains of increasing characteristic size in a variable one-species background up to $t \sim 10^3$. At larger $t$ the percolation between three-species spatial domains eventually leads to a population network of spiral patterns. The central panel of Fig. \ref{fig2}  depicts the evolution of the density of the different species and empty sites ($\rho_i$ and $\rho_0$, respectively). As in Fig. \ref{fig1}, the red, blue and green lines represent the densities of species 1, 2 and 3, respectively, but in this case there are also empty sites whose density is given by the grey bottom line. Fig. \ref{fig2} shows that in a May-Leonard implementation there is also a transient initial stage prior to an asymptotic regime in which the densities of the three species quickly approach nearly constant values, with $\rho_1 \gsim \rho_2>\rho_3$. However, the evolution is considerably slower and the fluctuations are considerably larger compared to a Lotka-Volterra implementation.

In the case of a May-Leonard implementation, the large coherent oscillations of the abundances of the various species in the early stages of simulations of the spatial RPS model with random initial conditions may result in a significant dependence of the surviving/most abundant species on the linear size of the lattice bellow a given linear size threshold $N_{th}$, even in the case of simulations with a large total simulation time. This is shown in Fig. \ref{fig3} which depicts the probability of single species survival and coexistence as a function of the linear size of the simulation lattice for a May-Leonard implementation of the spatial stochastic RPS model with $m=0.3$, $p=0.35$, $r=0.35$ and ${\mathcal P}_w=0.5$.  Each point was estimated from  $10^4$  simulations with a total simulation time equal to $2 \times 10^4$ generations, starting from random initial conditions with $\rho_1=\rho_2 = \rho_3 =
1/3$. The error bars are always much smaller than the size of the symbols: the one-sigma uncertainty in the value of $P$, at each point, may be estimated as $(P(1-P)/10^4)^{1/2}$, with a maximum of $5 \times 10^{-3}$ for $P=0.5$. Fig. \ref{fig3} shows that for linear sizes $N > N_{th} \sim  30$, the `weakest' species has the largest probability to survive, but this no longer holds for $N < N_{th}$. 

\begin{figure}[t]
	\centering
		\includegraphics{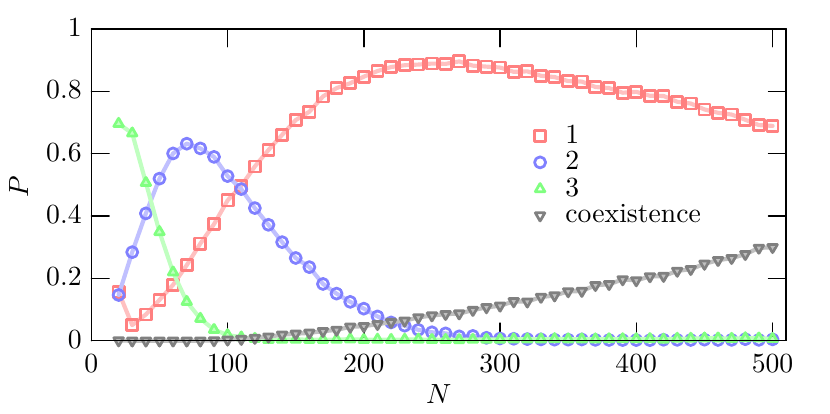}
	\caption{The same as in Fig. \ref{fig5} but for $m=0.5$, $p=0.25$, $r=0.25$, and ${\mathcal P}_w=0.5$.}
	\label{fig4}
\end{figure}

Figure \ref{fig4} is analogous to Fig. \ref{fig3} but considers a different choice of model parameters: $m=0.5$, $p=0.25$, $r=0.25$, and ${\mathcal P}_w=0.5$. The larger mobility leads to an increase of the lattice linear size above which the `weakest' is the most likely to survive (in this case, $N_{th} \sim  110$), thus showing that this threshold is strongly dependent on the choice of models parameters. We also verified that the use of a Moore neighbourhood --- composed of a central cell (the active one) and the eight cells that surround it --- leads to similar qualitative results to the ones presented in Figs. \ref{fig3} and \ref{fig4} for a von Neumann neighbourhood, albeit with significantly larger linear thresholds ($N_{th} \sim 70 $ and $N_{th} \sim 370$, respectively). Hence, the small linear size ($N=50$) associated to the limited total simulation time ($t=250$) of the simulations performed in Ref.  \cite{2019-Menezes-EPL-126-18003} using a Moore neighbourhood explains the reported impact of some of the model parameters on the determination of the surviving species in a May-Leonard implementation of the RPS model.

\begin{figure}[t]
	\centering
		\includegraphics{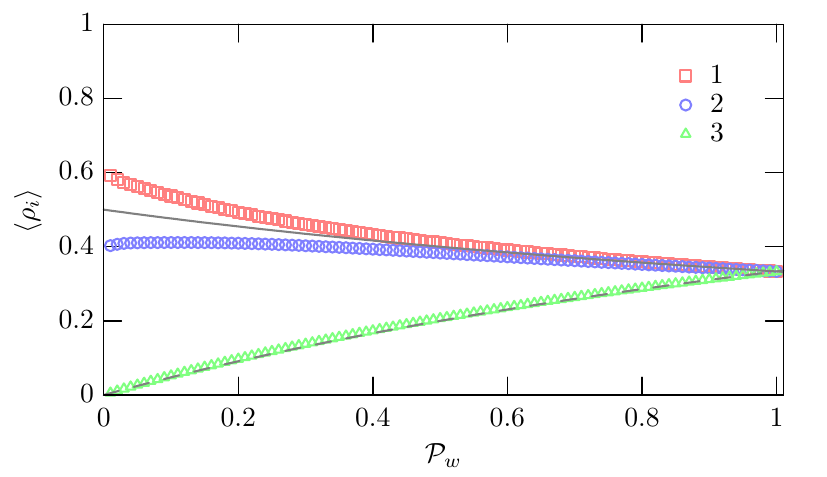}
	\caption{The value of the average density as a function of $\mathcal{P}_w$ for a Lotka-Volterra implementation of the spatial stochastic RPS model with $m=0.5$, and  $p=0.5$. Each point results from an average over the last $10^4$ generations of $1000^2$ simulations with a timespan equal to $1.5 \times 10^3$ generations. The lines represent the stationary solution, given in Eq. (\ref{rho123LV}), for the density of the species $1$ and $2$ (solid line) and $3$ (dashed line) obtained in the context of a Lotka-Volterra implementation of the RPS model.}
	\label{fig5}
\end{figure}

Figure \ref{fig5} shows the value of the average density of the three species as a function of $\mathcal{P}_w$ for a Lotka-Volterra implementation of the spatial stochastic RPS model. The data points result from an average over the last $10^4$ generations of simulations with a timespan equal to $1.5 \times 10^3$ generations performed on a $1000^2$ lattice. The results for ${\mathcal P}_w=1$ were computed first, starting from random initial conditions (as in Fig. \ref{fig1}). The final conditions of each simulation with ${\mathcal P}_w=1$ were used as initial conditions for a new simulation with ${\mathcal P}_w=1-0.01$. This procedure was repeated until ${\mathcal P}_w=0.01$ was reached. Such an approach was used in order to allow for a fast convergence (we verified that, with such conditions, $5 \times 10^3$ simulations are sufficient for $\langle \rho_i \rangle$ to attain its asymptotic value). In this way the large oscillations at the initial stages of simulations with random initial conditions shown in Figs. \ref{fig1} and \ref{fig2} --- which, depending on the value of ${\mathcal P}_w$ could be responsible for the loss of coexistence on a relatively short timescale --- are avoided.  Hence, this choice of initial conditions allowed us to obtain results which, in the case random initial conditions, would require larger simulation lattices. Figure \ref{fig5} shows that the `weakest' species is always the most abundant, thus having a competitive advantage over the others, specially over its predator. Figure \ref{fig5} also shows that competitive advantage over the other species increases as ${\mathcal P}_w$ decreases --- the `weakest' species and its prey having similar abundances for ${\mathcal P}_w > 0.6$. The lines in Fig. \ref{fig5} represent the stationary solution, given in Eq. (\ref{rho123LV}), for the density of the species $1$ and $2$ (solid line) and $3$ (dashed line) obtained in the context of a Lotka-Volterra implementation of the non-spatial RPS model. Notice the remarkable agreement between the spatial and non-spatial results in a Lotka-Volterra implementation of the RPS model.

\begin{figure}[t]
	\centering
		\includegraphics{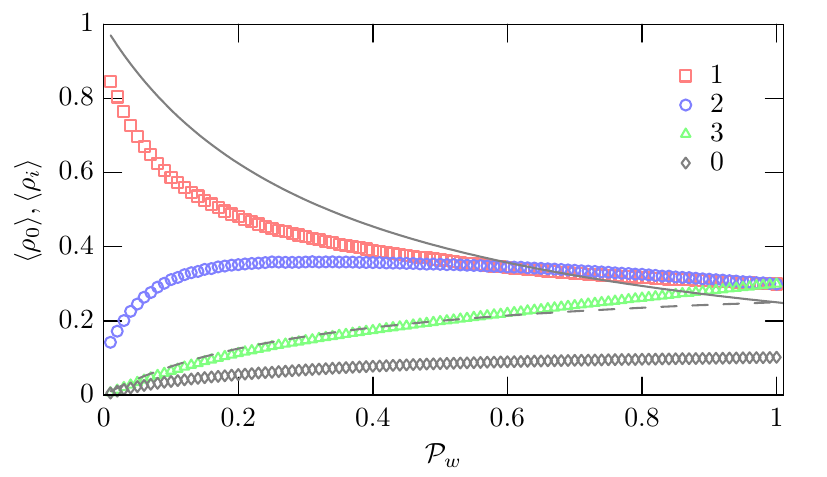}
	\caption{Same as in Fig. \ref{fig5}, but for a May-Leonard realization of the RPS model with $m=0.5$ and $p=r=0.25$. The lines represent the stationary solution, given in Eqs. (\ref{rho0123MLa}-\ref{rho0123MLc}), for the density of the species $1$ (solid line) and of species $2$ and $3$, and empty sites (dashed line) obtained in the context of a May-Leonard implementation of the RPS model with $p=r$.}
	\label{fig6}
\end{figure}

Figure \ref{fig6} is analogous to Fig. \ref{fig5}, except that, in this case, a May-Leonard realization of the RPS model with $m=0.5$, $p=r=0.25$ is considered.  Notice that, despite the considerably different population dynamics and spatial patterns, the late time asymptotic values of the relative abundances of the three species (as a function of ${\mathcal P}_w$) obtained for a May-Leonard implementation are qualitatively similar to the ones shown in Figure \ref{fig5} for a Lotka-Volterra implementation. In both cases the `weakest' species generally has a competitive advantage over the others --- specially over its predator. Again, this is particularly true at low values of ${\mathcal P}_w$ in both implementations. However, in a May-Leonard implementation there is a regime, for $0.6 <{\mathcal P}_w < 1$, in which the prey of the `weakest' species (species 2) is the dominant one, albeit only by a small margin. The lines in Fig. \ref{fig6} represent the stationary solution, given in Eqs. (\ref{rho0123MLa}-\ref{rho0123MLc}), for the density of the species $1$ (solid line) and of species $2$ and $3$, and empty sites (dashed line) obtained in the context of a May-Leonard implementation of the non-spatial RPS model with $r=p$. In the case of a May-Leonard implementation of the RPS model the differences between the spatial and non-spatial results are significant. This is a result of the distinct spatial structure and of the associated dynamics generated in a May-Leonard implementation of the spatial RPS model.

We verified that the results shown in Figs. \ref{fig5} and \ref{fig6} would remain essentially unchanged if a Moore neighbourhood had been used instead of a von Neumann one. We also checked that for other values of $p$, $m$, and $r$ (with $r=0$ in the case of a Lotka-Volterra implementation) the results obtained for the dependence of the values of the average densities on $\mathcal{P}_w$ are qualitatively similar to the ones shown in Figs. \ref{fig5} and \ref{fig6}. This is partially explained by the fact that different choices of the parameters may, to some extent, be absorbed by a redefinition of time and spatial units. In particular, the stationary solution for the values of the average densities in the non-spatial Lotka-Volterra implementation given in Eq. (\ref{rho123LV}) only depends on ${\mathcal P}_w$ --- any dependence on $p$ may be absorbed in the choice of a different time unit, which does not affect the stationary solutions. In the case of a May-Leonard implementation the stationary average densities given in Eqs. (\ref{rho0123MLa}-\ref{rho0123MLc}) depend both on ${\mathcal P}_w$ and $r/p$, but the dependence on $r/p$ has no impact on which species is the most abundant one. In a spatial version of the PRS model mobility also plays an important role. However, in a mean field description of a May-Leonard implementation of the RPS model changes of $m$ may be absorbed by an appropriate redefinition of spatial units \cite{Reichenbach-N-448-1046}.

\section{Conclusions \label{sec5}}

In this paper we revisited the problem of the predominance of the `weakest' species in the context of Lotka-Volterra and May-Leonard implementations of a spatial stochastic RPS model in which one of the species has a reduced predation probability. We have shown that, despite the significant dynamical differences between Lotka-Volterra and May-Leonard implementations of the RPS model, for sufficiently large lattices the late time values of the relative abundances of the various species display similar qualitative dependencies on the reduced predation probability (parameterized by ${\mathcal P}_w$) --- with the `weakest' species being the most abundant or having an average density extremely close to that of the most abundant species. We have also found that if the linear size of the lattice or the total simulation time  are not sufficiently large, then the probability of species survival is strongly dependent on initial conditions, in which case a higher probability of survival of the `weakest' species does not generally happen.

\begin{acknowledgments}
P.P.A. acknowledges the support by FCT/MCTES through Portuguese funds (PIDDAC) by the grant UID/FIS/04434/2019. B.F.O. and R.S.T. thank CAPES, Funda\c c\~ao Arauc\'aria, and INCT-FCx (CNPq/FAPESP) for financial and computational support.
\end{acknowledgments}


\begin{thebibliography}{46}
\expandafter\ifx\csname natexlab\endcsname\relax\def\natexlab#1{#1}\fi
\expandafter\ifx\csname bibnamefont\endcsname\relax
  \def\bibnamefont#1{#1}\fi
\expandafter\ifx\csname bibfnamefont\endcsname\relax
  \def\bibfnamefont#1{#1}\fi
\expandafter\ifx\csname citenamefont\endcsname\relax
  \def\citenamefont#1{#1}\fi
\expandafter\ifx\csname url\endcsname\relax
  \def\url#1{\texttt{#1}}\fi
\expandafter\ifx\csname urlprefix\endcsname\relax\def\urlprefix{URL }\fi
\providecommand{\bibinfo}[2]{#2}
\providecommand{\eprint}[2][]{\url{#2}}

\bibitem[{\citenamefont{Kerr et~al.}(2002)\citenamefont{Kerr, Riley, Feldman,
  and Bohannan}}]{2002-Kerr-N-418-171}
\bibinfo{author}{\bibfnamefont{B.}~\bibnamefont{Kerr}},
  \bibinfo{author}{\bibfnamefont{M.~A.} \bibnamefont{Riley}},
  \bibinfo{author}{\bibfnamefont{M.~W.} \bibnamefont{Feldman}},
  \bibnamefont{and} \bibinfo{author}{\bibfnamefont{B.~J.~M.}
  \bibnamefont{Bohannan}}, \bibinfo{journal}{Nature}
  \textbf{\bibinfo{volume}{418}}, \bibinfo{pages}{171} (\bibinfo{year}{2002}).

\bibitem[{\citenamefont{Reichenbach et~al.}(2007)\citenamefont{Reichenbach,
  Mobilia, and Frey}}]{Reichenbach-N-448-1046}
\bibinfo{author}{\bibfnamefont{T.}~\bibnamefont{Reichenbach}},
  \bibinfo{author}{\bibfnamefont{M.}~\bibnamefont{Mobilia}}, \bibnamefont{and}
  \bibinfo{author}{\bibfnamefont{E.}~\bibnamefont{Frey}},
  \bibinfo{journal}{Nature} \textbf{\bibinfo{volume}{448}},
  \bibinfo{pages}{1046} (\bibinfo{year}{2007}).

\bibitem[{\citenamefont{{Lotka}}(1920)}]{1920PNAS....6..410L}
\bibinfo{author}{\bibfnamefont{A.~J.} \bibnamefont{{Lotka}}},
  \bibinfo{journal}{Proceedings of the National Academy of Science}
  \textbf{\bibinfo{volume}{6}}, \bibinfo{pages}{410} (\bibinfo{year}{1920}).

\bibitem[{\citenamefont{{Volterra}}(1926)}]{1926Natur.118..558V}
\bibinfo{author}{\bibfnamefont{V.}~\bibnamefont{{Volterra}}},
  \bibinfo{journal}{\nat} \textbf{\bibinfo{volume}{118}}, \bibinfo{pages}{558}
  (\bibinfo{year}{1926}).

\bibitem[{\citenamefont{May and Leonard}(1975)}]{May-Leonard}
\bibinfo{author}{\bibfnamefont{R.}~\bibnamefont{May}} \bibnamefont{and}
  \bibinfo{author}{\bibfnamefont{W.}~\bibnamefont{Leonard}},
  \bibinfo{journal}{SIAM Journal on Applied Mathematics}
  \textbf{\bibinfo{volume}{29}}, \bibinfo{pages}{243} (\bibinfo{year}{1975}).

\bibitem[{\citenamefont{Sinervo and Lively}(1996)}]{lizards}
\bibinfo{author}{\bibfnamefont{B.}~\bibnamefont{Sinervo}} \bibnamefont{and}
  \bibinfo{author}{\bibfnamefont{C.~M.} \bibnamefont{Lively}},
  \bibinfo{journal}{Nature} \textbf{\bibinfo{volume}{380}},
  \bibinfo{pages}{240} (\bibinfo{year}{1996}).

\bibitem[{\citenamefont{Kirkup and Riley}(2004)}]{bacteria}
\bibinfo{author}{\bibfnamefont{B.~C.} \bibnamefont{Kirkup}} \bibnamefont{and}
  \bibinfo{author}{\bibfnamefont{M.~A.} \bibnamefont{Riley}},
  \bibinfo{journal}{Nature} \textbf{\bibinfo{volume}{428}},
  \bibinfo{pages}{412} (\bibinfo{year}{2004}).

\bibitem[{\citenamefont{Szab{\'{o}} et~al.}(2004)\citenamefont{Szab{\'{o}},
  Szolnoki, and Izs{\'{a}}k}}]{2004-Szabo-JPAMG-31-2599}
\bibinfo{author}{\bibfnamefont{G.}~\bibnamefont{Szab{\'{o}}}},
  \bibinfo{author}{\bibfnamefont{A.}~\bibnamefont{Szolnoki}}, \bibnamefont{and}
  \bibinfo{author}{\bibfnamefont{R.}~\bibnamefont{Izs{\'{a}}k}},
  \bibinfo{journal}{Journal of Physics A: Mathematical and General}
  \textbf{\bibinfo{volume}{37}}, \bibinfo{pages}{2599} (\bibinfo{year}{2004}).

\bibitem[{\citenamefont{Zhang et~al.}(2009)\citenamefont{Zhang, Chen, Qi, and
  Qing}}]{2009-Zhang-PRE-79-062901}
\bibinfo{author}{\bibfnamefont{G.-Y.} \bibnamefont{Zhang}},
  \bibinfo{author}{\bibfnamefont{Y.}~\bibnamefont{Chen}},
  \bibinfo{author}{\bibfnamefont{W.-K.} \bibnamefont{Qi}}, \bibnamefont{and}
  \bibinfo{author}{\bibfnamefont{S.-M.} \bibnamefont{Qing}},
  \bibinfo{journal}{Phys. Rev. E} \textbf{\bibinfo{volume}{79}},
  \bibinfo{pages}{062901} (\bibinfo{year}{2009}).

\bibitem[{\citenamefont{Laird}(2014)}]{2014-Laird-Oikos-123-472}
\bibinfo{author}{\bibfnamefont{R.~A.} \bibnamefont{Laird}},
  \bibinfo{journal}{Oikos} \textbf{\bibinfo{volume}{123}}, \bibinfo{pages}{472}
  (\bibinfo{year}{2014}).

\bibitem[{\citenamefont{Rulquin and
  Arenzon}(2014)}]{2014-Rulquin-PRE-89-032133}
\bibinfo{author}{\bibfnamefont{C.}~\bibnamefont{Rulquin}} \bibnamefont{and}
  \bibinfo{author}{\bibfnamefont{J.~J.} \bibnamefont{Arenzon}},
  \bibinfo{journal}{Phys. Rev. E} \textbf{\bibinfo{volume}{89}},
  \bibinfo{pages}{032133} (\bibinfo{year}{2014}).

\bibitem[{\citenamefont{Peltom\"{a}ki and
  Alava}(2008)}]{2008-Peltomaki-PRE-78-031906}
\bibinfo{author}{\bibfnamefont{M.}~\bibnamefont{Peltom\"{a}ki}}
  \bibnamefont{and} \bibinfo{author}{\bibfnamefont{M.}~\bibnamefont{Alava}},
  \bibinfo{journal}{Phys. Rev. E} \textbf{\bibinfo{volume}{78}},
  \bibinfo{pages}{031906} (\bibinfo{year}{2008}).

\bibitem[{\citenamefont{Szab{\'{o}} et~al.}(2008)\citenamefont{Szab{\'{o}},
  Szolnoki, and Borsos}}]{2008-Szabo-PRE-77-041919}
\bibinfo{author}{\bibfnamefont{G.}~\bibnamefont{Szab{\'{o}}}},
  \bibinfo{author}{\bibfnamefont{A.}~\bibnamefont{Szolnoki}}, \bibnamefont{and}
  \bibinfo{author}{\bibfnamefont{I.}~\bibnamefont{Borsos}},
  \bibinfo{journal}{Phys. Rev. E} \textbf{\bibinfo{volume}{77}},
  \bibinfo{pages}{041919} (\bibinfo{year}{2008}).

\bibitem[{\citenamefont{Allesina and
  Levine}(2011)}]{2011-Allesina-PNAS-108-5638}
\bibinfo{author}{\bibfnamefont{S.}~\bibnamefont{Allesina}} \bibnamefont{and}
  \bibinfo{author}{\bibfnamefont{J.~M.} \bibnamefont{Levine}},
  \bibinfo{journal}{PNAS} \textbf{\bibinfo{volume}{108}}, \bibinfo{pages}{5638}
  (\bibinfo{year}{2011}).

\bibitem[{\citenamefont{Avelino
  et~al.}(2012{\natexlab{a}})\citenamefont{Avelino, Bazeia, Losano, and
  Menezes}}]{2012-Avelino-PRE-86-031119}
\bibinfo{author}{\bibfnamefont{P.~P.} \bibnamefont{Avelino}},
  \bibinfo{author}{\bibfnamefont{D.}~\bibnamefont{Bazeia}},
  \bibinfo{author}{\bibfnamefont{L.}~\bibnamefont{Losano}}, \bibnamefont{and}
  \bibinfo{author}{\bibfnamefont{J.}~\bibnamefont{Menezes}},
  \bibinfo{journal}{Phys. Rev. E} \textbf{\bibinfo{volume}{86}},
  \bibinfo{pages}{031119} (\bibinfo{year}{2012}{\natexlab{a}}).

\bibitem[{\citenamefont{Avelino
  et~al.}(2012{\natexlab{b}})\citenamefont{Avelino, Bazeia, Losano, Menezes,
  and Oliveira}}]{2012-Avelino-PRE-86-036112}
\bibinfo{author}{\bibfnamefont{P.~P.} \bibnamefont{Avelino}},
  \bibinfo{author}{\bibfnamefont{D.}~\bibnamefont{Bazeia}},
  \bibinfo{author}{\bibfnamefont{L.}~\bibnamefont{Losano}},
  \bibinfo{author}{\bibfnamefont{J.}~\bibnamefont{Menezes}}, \bibnamefont{and}
  \bibinfo{author}{\bibfnamefont{B.~F.} \bibnamefont{Oliveira}},
  \bibinfo{journal}{Phys. Rev. E} \textbf{\bibinfo{volume}{86}},
  \bibinfo{pages}{036112} (\bibinfo{year}{2012}{\natexlab{b}}).

\bibitem[{\citenamefont{Li et~al.}(2012)\citenamefont{Li, Dong, and
  Yang}}]{2012-Li-PA-391-125}
\bibinfo{author}{\bibfnamefont{Y.}~\bibnamefont{Li}},
  \bibinfo{author}{\bibfnamefont{L.}~\bibnamefont{Dong}}, \bibnamefont{and}
  \bibinfo{author}{\bibfnamefont{G.}~\bibnamefont{Yang}},
  \bibinfo{journal}{Physica A: Statistical Mechanics and its Applications}
  \textbf{\bibinfo{volume}{391}}, \bibinfo{pages}{125} (\bibinfo{year}{2012}).

\bibitem[{\citenamefont{Roman et~al.}(2012)\citenamefont{Roman, Konrad, and
  Pleimling}}]{2012-Roman-JSMTE-2012-p07014}
\bibinfo{author}{\bibfnamefont{A.}~\bibnamefont{Roman}},
  \bibinfo{author}{\bibfnamefont{D.}~\bibnamefont{Konrad}}, \bibnamefont{and}
  \bibinfo{author}{\bibfnamefont{M.}~\bibnamefont{Pleimling}},
  \bibinfo{journal}{Journal of Statistical Mechanics: Theory and Experiment}
  \textbf{\bibinfo{volume}{2012}}, \bibinfo{pages}{P07014}
  (\bibinfo{year}{2012}).

\bibitem[{\citenamefont{L\"{u}tz et~al.}(2013)\citenamefont{L\"{u}tz,
  Risau-Gusman, and Arenzon}}]{2013-Lutz-JTB-317-286}
\bibinfo{author}{\bibfnamefont{A.~F.} \bibnamefont{L\"{u}tz}},
  \bibinfo{author}{\bibfnamefont{S.}~\bibnamefont{Risau-Gusman}},
  \bibnamefont{and} \bibinfo{author}{\bibfnamefont{J.~J.}
  \bibnamefont{Arenzon}}, \bibinfo{journal}{Journal of Theoretical Biology}
  \textbf{\bibinfo{volume}{317}}, \bibinfo{pages}{286} (\bibinfo{year}{2013}).

\bibitem[{\citenamefont{Roman et~al.}(2013)\citenamefont{Roman, Dasgupta, and
  Pleimling}}]{2013-Roman-PRE-87-032148}
\bibinfo{author}{\bibfnamefont{A.}~\bibnamefont{Roman}},
  \bibinfo{author}{\bibfnamefont{D.}~\bibnamefont{Dasgupta}}, \bibnamefont{and}
  \bibinfo{author}{\bibfnamefont{M.}~\bibnamefont{Pleimling}},
  \bibinfo{journal}{Phys. Rev. E} \textbf{\bibinfo{volume}{87}},
  \bibinfo{pages}{032148} (\bibinfo{year}{2013}).

\bibitem[{\citenamefont{Cheng et~al.}(2014)\citenamefont{Cheng, Yao, Huang,
  Park, Do, and Lai}}]{2014-Cheng-SR-4-7486}
\bibinfo{author}{\bibfnamefont{H.}~\bibnamefont{Cheng}},
  \bibinfo{author}{\bibfnamefont{N.}~\bibnamefont{Yao}},
  \bibinfo{author}{\bibfnamefont{Z.-G.} \bibnamefont{Huang}},
  \bibinfo{author}{\bibfnamefont{J.}~\bibnamefont{Park}},
  \bibinfo{author}{\bibfnamefont{Y.}~\bibnamefont{Do}}, \bibnamefont{and}
  \bibinfo{author}{\bibfnamefont{Y.-C.} \bibnamefont{Lai}},
  \bibinfo{journal}{Scientific Reports} \textbf{\bibinfo{volume}{4}},
  \bibinfo{pages}{7486} (\bibinfo{year}{2014}).

\bibitem[{\citenamefont{Szolnoki et~al.}(2014)\citenamefont{Szolnoki, Mobilia,
  Jiang, Szczesny, Rucklidge, and Perc}}]{2014-Szolnoki-JRSI-11-0735}
\bibinfo{author}{\bibfnamefont{A.}~\bibnamefont{Szolnoki}},
  \bibinfo{author}{\bibfnamefont{M.}~\bibnamefont{Mobilia}},
  \bibinfo{author}{\bibfnamefont{L.-L.} \bibnamefont{Jiang}},
  \bibinfo{author}{\bibfnamefont{B.}~\bibnamefont{Szczesny}},
  \bibinfo{author}{\bibfnamefont{A.~M.} \bibnamefont{Rucklidge}},
  \bibnamefont{and} \bibinfo{author}{\bibfnamefont{M.}~\bibnamefont{Perc}},
  \bibinfo{journal}{Journal of The Royal Society Interface}
  \textbf{\bibinfo{volume}{11}}, \bibinfo{pages}{20140735}
  (\bibinfo{year}{2014}).

\bibitem[{\citenamefont{Kang et~al.}(2016)\citenamefont{Kang, Pan, Wang, and
  He}}]{2016-Kang-Entropy-18-284}
\bibinfo{author}{\bibfnamefont{Y.}~\bibnamefont{Kang}},
  \bibinfo{author}{\bibfnamefont{Q.}~\bibnamefont{Pan}},
  \bibinfo{author}{\bibfnamefont{X.}~\bibnamefont{Wang}}, \bibnamefont{and}
  \bibinfo{author}{\bibfnamefont{M.}~\bibnamefont{He}},
  \bibinfo{journal}{Entropy} \textbf{\bibinfo{volume}{18}},
  \bibinfo{pages}{284} (\bibinfo{year}{2016}).

\bibitem[{\citenamefont{Roman et~al.}(2016)\citenamefont{Roman, Dasgupta, and
  Pleimling}}]{2016-Roman-JTB-403-10}
\bibinfo{author}{\bibfnamefont{A.}~\bibnamefont{Roman}},
  \bibinfo{author}{\bibfnamefont{D.}~\bibnamefont{Dasgupta}}, \bibnamefont{and}
  \bibinfo{author}{\bibfnamefont{M.}~\bibnamefont{Pleimling}},
  \bibinfo{journal}{Journal of Theoretical Biology}
  \textbf{\bibinfo{volume}{403}}, \bibinfo{pages}{10} (\bibinfo{year}{2016}).

\bibitem[{\citenamefont{Brown and Pleimling}(2017)}]{2017-Brown-PRE-96-012147}
\bibinfo{author}{\bibfnamefont{B.~L.} \bibnamefont{Brown}} \bibnamefont{and}
  \bibinfo{author}{\bibfnamefont{M.}~\bibnamefont{Pleimling}},
  \bibinfo{journal}{Phys. Rev. E} \textbf{\bibinfo{volume}{96}},
  \bibinfo{pages}{012147} (\bibinfo{year}{2017}).

\bibitem[{\citenamefont{Park et~al.}(2017)\citenamefont{Park, Do, Jang, and
  Lai}}]{2017-Park-SR-7-7465}
\bibinfo{author}{\bibfnamefont{J.}~\bibnamefont{Park}},
  \bibinfo{author}{\bibfnamefont{Y.}~\bibnamefont{Do}},
  \bibinfo{author}{\bibfnamefont{B.}~\bibnamefont{Jang}}, \bibnamefont{and}
  \bibinfo{author}{\bibfnamefont{Y.-C.} \bibnamefont{Lai}},
  \bibinfo{journal}{Scientific Reports} \textbf{\bibinfo{volume}{7}},
  \bibinfo{pages}{7465} (\bibinfo{year}{2017}).

\bibitem[{\citenamefont{Bazeia et~al.}(2017)\citenamefont{Bazeia, Menezes,
  de~Oliveira, and Ramos}}]{2017-Bazeia-EPL-119-58003}
\bibinfo{author}{\bibfnamefont{D.}~\bibnamefont{Bazeia}},
  \bibinfo{author}{\bibfnamefont{J.}~\bibnamefont{Menezes}},
  \bibinfo{author}{\bibfnamefont{B.~F.} \bibnamefont{de~Oliveira}},
  \bibnamefont{and} \bibinfo{author}{\bibfnamefont{J.~G. G.~S.}
  \bibnamefont{Ramos}}, \bibinfo{journal}{EPL} \textbf{\bibinfo{volume}{119}},
  \bibinfo{pages}{58003} (\bibinfo{year}{2017}).

\bibitem[{\citenamefont{Souza-Filho et~al.}(2017)\citenamefont{Souza-Filho,
  Bazeia, and Ramos}}]{2017-Souza-Filho-PRE-95-062411}
\bibinfo{author}{\bibfnamefont{C.~A.} \bibnamefont{Souza-Filho}},
  \bibinfo{author}{\bibfnamefont{D.}~\bibnamefont{Bazeia}}, \bibnamefont{and}
  \bibinfo{author}{\bibfnamefont{J.~G. G.~S.} \bibnamefont{Ramos}},
  \bibinfo{journal}{Phys. Rev. E} \textbf{\bibinfo{volume}{95}},
  \bibinfo{pages}{062411} (\bibinfo{year}{2017}).

\bibitem[{\citenamefont{Esmaeili et~al.}(2018)\citenamefont{Esmaeili, Brown,
  and Pleimling}}]{2018-Shadisadt-PRE-98-062105}
\bibinfo{author}{\bibfnamefont{S.}~\bibnamefont{Esmaeili}},
  \bibinfo{author}{\bibfnamefont{B.~L.} \bibnamefont{Brown}}, \bibnamefont{and}
  \bibinfo{author}{\bibfnamefont{M.}~\bibnamefont{Pleimling}},
  \bibinfo{journal}{Phys. Rev. E} \textbf{\bibinfo{volume}{98}},
  \bibinfo{pages}{062105} (\bibinfo{year}{2018}).

\bibitem[{\citenamefont{Avelino et~al.}(2019)\citenamefont{Avelino, Menezes,
  de~Oliveira, and Pereira}}]{2019-Avelino-PRE-99-052310}
\bibinfo{author}{\bibfnamefont{P.~P.} \bibnamefont{Avelino}},
  \bibinfo{author}{\bibfnamefont{J.}~\bibnamefont{Menezes}},
  \bibinfo{author}{\bibfnamefont{B.~F.} \bibnamefont{de~Oliveira}},
  \bibnamefont{and} \bibinfo{author}{\bibfnamefont{T.~A.}
  \bibnamefont{Pereira}}, \bibinfo{journal}{Phys. Rev. E}
  \textbf{\bibinfo{volume}{99}}, \bibinfo{pages}{052310}
  (\bibinfo{year}{2019}).

\bibitem[{\citenamefont{Bazeia et~al.}(2019)\citenamefont{Bazeia, de~Oliveira,
  and Szolnoki}}]{2019-Bazeia-PRE-99-052408}
\bibinfo{author}{\bibfnamefont{D.}~\bibnamefont{Bazeia}},
  \bibinfo{author}{\bibfnamefont{B.~F.} \bibnamefont{de~Oliveira}},
  \bibnamefont{and} \bibinfo{author}{\bibfnamefont{A.}~\bibnamefont{Szolnoki}},
  \bibinfo{journal}{Phys. Rev. E} \textbf{\bibinfo{volume}{99}},
  \bibinfo{pages}{052408} (\bibinfo{year}{2019}).

\bibitem[{\citenamefont{Avelino
  et~al.}(2014{\natexlab{a}})\citenamefont{Avelino, Bazeia, Losano, Menezes,
  and de~Oliveira}}]{2014-Avelino-PRE-89-042710}
\bibinfo{author}{\bibfnamefont{P.~P.} \bibnamefont{Avelino}},
  \bibinfo{author}{\bibfnamefont{D.}~\bibnamefont{Bazeia}},
  \bibinfo{author}{\bibfnamefont{L.}~\bibnamefont{Losano}},
  \bibinfo{author}{\bibfnamefont{J.}~\bibnamefont{Menezes}}, \bibnamefont{and}
  \bibinfo{author}{\bibfnamefont{B.~F.} \bibnamefont{de~Oliveira}},
  \bibinfo{journal}{Phys. Rev. E} \textbf{\bibinfo{volume}{89}},
  \bibinfo{pages}{042710} (\bibinfo{year}{2014}{\natexlab{a}}).

\bibitem[{\citenamefont{Avelino
  et~al.}(2014{\natexlab{b}})\citenamefont{Avelino, Bazeia, Menezes, and
  de~Oliveira}}]{2014-Avelino-PLA-378-393}
\bibinfo{author}{\bibfnamefont{P.~P.} \bibnamefont{Avelino}},
  \bibinfo{author}{\bibfnamefont{D.}~\bibnamefont{Bazeia}},
  \bibinfo{author}{\bibfnamefont{J.}~\bibnamefont{Menezes}}, \bibnamefont{and}
  \bibinfo{author}{\bibfnamefont{B.~F.} \bibnamefont{de~Oliveira}},
  \bibinfo{journal}{Physics Letters A} \textbf{\bibinfo{volume}{378}},
  \bibinfo{pages}{393} (\bibinfo{year}{2014}{\natexlab{b}}).

\bibitem[{\citenamefont{Avelino et~al.}(2017)\citenamefont{Avelino, Bazeia,
  Losano, Menezes, and de~Oliveira}}]{2017-Avelino-PLA-381-1014}
\bibinfo{author}{\bibfnamefont{P.~P.} \bibnamefont{Avelino}},
  \bibinfo{author}{\bibfnamefont{D.}~\bibnamefont{Bazeia}},
  \bibinfo{author}{\bibfnamefont{L.}~\bibnamefont{Losano}},
  \bibinfo{author}{\bibfnamefont{J.}~\bibnamefont{Menezes}}, \bibnamefont{and}
  \bibinfo{author}{\bibfnamefont{B.~F.} \bibnamefont{de~Oliveira}},
  \bibinfo{journal}{Physics Letters A} \textbf{\bibinfo{volume}{381}},
  \bibinfo{pages}{1014 } (\bibinfo{year}{2017}).

\bibitem[{\citenamefont{Szab\'o and
  Cz\'ar\'an}(2001)}]{2001-Szabo-PRE-63-061904}
\bibinfo{author}{\bibfnamefont{G.}~\bibnamefont{Szab\'o}} \bibnamefont{and}
  \bibinfo{author}{\bibfnamefont{T.}~\bibnamefont{Cz\'ar\'an}},
  \bibinfo{journal}{Phys. Rev. E} \textbf{\bibinfo{volume}{63}},
  \bibinfo{pages}{061904} (\bibinfo{year}{2001}).

\bibitem[{\citenamefont{Szab\'o and
  Arial~Sznaider}(2004)}]{2004-Szabo-PRE-69-031911}
\bibinfo{author}{\bibfnamefont{G.}~\bibnamefont{Szab\'o}} \bibnamefont{and}
  \bibinfo{author}{\bibfnamefont{G.}~\bibnamefont{Arial~Sznaider}},
  \bibinfo{journal}{Phys. Rev. E} \textbf{\bibinfo{volume}{69}},
  \bibinfo{pages}{031911} (\bibinfo{year}{2004}).

\bibitem[{\citenamefont{Szolnoki and
  Szab{\'{o}}}(2004)}]{2004-Szolnoki-PRE-70-037102}
\bibinfo{author}{\bibfnamefont{A.}~\bibnamefont{Szolnoki}} \bibnamefont{and}
  \bibinfo{author}{\bibfnamefont{G.}~\bibnamefont{Szab{\'{o}}}},
  \bibinfo{journal}{Phys. Rev. E} \textbf{\bibinfo{volume}{70}},
  \bibinfo{pages}{037102} (\bibinfo{year}{2004}).

\bibitem[{\citenamefont{Perc et~al.}(2007)\citenamefont{Perc, Szolnoki, and
  Szab\'o}}]{2007-Perc-PRE-75-052102}
\bibinfo{author}{\bibfnamefont{M.}~\bibnamefont{Perc}},
  \bibinfo{author}{\bibfnamefont{A.}~\bibnamefont{Szolnoki}}, \bibnamefont{and}
  \bibinfo{author}{\bibfnamefont{G.}~\bibnamefont{Szab\'o}},
  \bibinfo{journal}{Phys. Rev. E} \textbf{\bibinfo{volume}{75}},
  \bibinfo{pages}{052102} (\bibinfo{year}{2007}).

\bibitem[{\citenamefont{Szab\'o et~al.}(2007)\citenamefont{Szab\'o, Szolnoki,
  and Sznaider}}]{2007-Szabo-PRE-76-051921}
\bibinfo{author}{\bibfnamefont{G.}~\bibnamefont{Szab\'o}},
  \bibinfo{author}{\bibfnamefont{A.}~\bibnamefont{Szolnoki}}, \bibnamefont{and}
  \bibinfo{author}{\bibfnamefont{G.~A.} \bibnamefont{Sznaider}},
  \bibinfo{journal}{Phys. Rev. E} \textbf{\bibinfo{volume}{76}},
  \bibinfo{pages}{051921} (\bibinfo{year}{2007}).

\bibitem[{\citenamefont{Szab{\'{o}} and
  Szolnoki}(2008)}]{2008-Szabo-PRE-77-011906}
\bibinfo{author}{\bibfnamefont{G.}~\bibnamefont{Szab{\'{o}}}} \bibnamefont{and}
  \bibinfo{author}{\bibfnamefont{A.}~\bibnamefont{Szolnoki}},
  \bibinfo{journal}{Phys. Rev. E} \textbf{\bibinfo{volume}{77}},
  \bibinfo{pages}{011906} (\bibinfo{year}{2008}).

\bibitem[{\citenamefont{Szolnoki et~al.}(2011)\citenamefont{Szolnoki,
  Szab{\'{o}}, and Czak{\'{o}}}}]{2011-Szolnoki-PRE-84-046106}
\bibinfo{author}{\bibfnamefont{A.}~\bibnamefont{Szolnoki}},
  \bibinfo{author}{\bibfnamefont{G.}~\bibnamefont{Szab{\'{o}}}},
  \bibnamefont{and}
  \bibinfo{author}{\bibfnamefont{L.}~\bibnamefont{Czak{\'{o}}}},
  \bibinfo{journal}{Phys. Rev. E} \textbf{\bibinfo{volume}{84}},
  \bibinfo{pages}{046106} (\bibinfo{year}{2011}).

\bibitem[{\citenamefont{Vukov et~al.}(2013)\citenamefont{Vukov, Szolnoki, and
  Szab{\'{o}}}}]{2013-Vukov-PRE-88-022123}
\bibinfo{author}{\bibfnamefont{J.}~\bibnamefont{Vukov}},
  \bibinfo{author}{\bibfnamefont{A.}~\bibnamefont{Szolnoki}}, \bibnamefont{and}
  \bibinfo{author}{\bibfnamefont{G.}~\bibnamefont{Szab{\'{o}}}},
  \bibinfo{journal}{Phys. Rev. E} \textbf{\bibinfo{volume}{88}},
  \bibinfo{pages}{022123} (\bibinfo{year}{2013}).

\bibitem[{\citenamefont{Bazeia et~al.}(2018)\citenamefont{Bazeia, de~Oliveira,
  and Szolnoki}}]{2018-Bazeia-EPL-124-68001}
\bibinfo{author}{\bibfnamefont{D.}~\bibnamefont{Bazeia}},
  \bibinfo{author}{\bibfnamefont{B.~F.} \bibnamefont{de~Oliveira}},
  \bibnamefont{and} \bibinfo{author}{\bibfnamefont{A.}~\bibnamefont{Szolnoki}},
  \bibinfo{journal}{{EPL} (Europhysics Letters)}
  \textbf{\bibinfo{volume}{124}}, \bibinfo{pages}{68001}
  (\bibinfo{year}{2018}).

\bibitem[{\citenamefont{Frean and Abraham}(2001)}]{2001-Frean-PRSLB-268-1323}
\bibinfo{author}{\bibfnamefont{M.}~\bibnamefont{Frean}} \bibnamefont{and}
  \bibinfo{author}{\bibfnamefont{E.~R.} \bibnamefont{Abraham}},
  \bibinfo{journal}{Proc. R. Soc. Lond. B} \textbf{\bibinfo{volume}{268}},
  \bibinfo{pages}{1323} (\bibinfo{year}{2001}).

\bibitem[{\citenamefont{{Berr} et~al.}(2009)\citenamefont{{Berr},
  {Reichenbach}, {Schottenloher}, and {Frey}}}]{2009PhRvL.102d8102B}
\bibinfo{author}{\bibfnamefont{M.}~\bibnamefont{{Berr}}},
  \bibinfo{author}{\bibfnamefont{T.}~\bibnamefont{{Reichenbach}}},
  \bibinfo{author}{\bibfnamefont{M.}~\bibnamefont{{Schottenloher}}},
  \bibnamefont{and} \bibinfo{author}{\bibfnamefont{E.}~\bibnamefont{{Frey}}},
  \bibinfo{journal}{Phys. Rev. Lett.} \textbf{\bibinfo{volume}{102}},
  \bibinfo{eid}{048102} (\bibinfo{year}{2009}).

\bibitem[{\citenamefont{Menezes et~al.}(2019)\citenamefont{Menezes, Moura, and
  Pereira}}]{2019-Menezes-EPL-126-18003}
\bibinfo{author}{\bibfnamefont{J.}~\bibnamefont{Menezes}},
  \bibinfo{author}{\bibfnamefont{B.}~\bibnamefont{Moura}}, \bibnamefont{and}
  \bibinfo{author}{\bibfnamefont{T.~A.} \bibnamefont{Pereira}},
  \bibinfo{journal}{{EPL} (Europhysics Letters)}
  \textbf{\bibinfo{volume}{126}}, \bibinfo{pages}{18003}
  (\bibinfo{year}{2019}).

\end{thebibliography}
\end{document}